\documentclass[12pt,reqno,fleqn,a4paper]{amsart}
\usepackage{mathrsfs}
\usepackage{bbm}
\usepackage{amsmath,amssymb}
\usepackage{color}

\usepackage[titletoc]{appendix}

\allowdisplaybreaks
 \newtheorem{thm}{Theorem}[section]
 
 \newtheorem{lem}[thm]{Lemma}
 \newtheorem{prp}[thm]{Proposition}

\newenvironment{prf}{\noindent {\it Proof} \ }{\hfill $\Box$}

\newcommand{\eqa}{\begin{eqnarray}}
\newcommand{\eeqa}{\end{eqnarray}}

\newcommand{\beq}{\begin{equation}}
\newcommand{\eeq}{\end{equation}}

\newcommand\pd{\partial}

\newcommand{\Gm}{\Gamma}

\newcommand\Z{\mathbb{Z}}
\newcommand\Zop{\mathbb{Z^{\mathrm{odd}}_+}}

\newcommand\B{\mathcal{B}}

\newcommand\cL{\mathcal{L}}
\newcommand\cM{\mathcal{M}}

\begin{document}
\title[{\small Block symmetry of Drinfeld-Sokolov hierarchy}]{\small Block (or Hamiltonian) Lie symmetry of dispersionless D type Drinfeld-Sokolov hierarchy}
\author[Chuanzhong Li,\ Jingsong He,\ Yucai Su]{
Chuanzhong Li$^*$,\  \ Jingsong He$^*$,\ \ Yucai Su$^\dag$}
\allowdisplaybreaks
\dedicatory {\small $^{*\,}$Department of Mathematics,  Ningbo University, Ningbo, 315211, China\\
$^{\dag\,}$Department of Mathematics, Tongji University, Shanghai 200092, China\\
lichuanzhong@nbu.edu.cn, hejingsong@nbu.edu.cn,  ycsu@tongji.edu.cn}
\thanks{Correspondence: hejingsong@nbu.edu.cn}

\date{}

\begin{abstract}
In this paper,  the dispersionless D type Drinfeld-Sokolov hierar-
chy, i.e. a reduction of the dispersionless two-component BKP hierarchy,
is studied.   The  additional symmetry flows of this herarchy are presented.
These flows form an infinite dimensional Lie algebra of Block type as well as a Lie algebra of Hamiltonian type.
\end{abstract}
\allowdisplaybreaks
\maketitle{\allowdisplaybreaks}
{\small \noindent {\bf PACS numbers:} 02.20.Sv, 02.20.Tw, 02.30.Ik

}
\allowdisplaybreaks
\setcounter{section}{0}\section{Introduction}

Additional symmetry is an  interesting  topic in the study of  integrable hierarchies which has been studied extensively in recent years.
Additional symmetries of the Kadomtsev-Petviashvili (KP) hierarchy
introduced  by Orlov and Shulman \cite{os1} contain one
kind of important symmetry, namely, the Virasoro symmetry, which is closely
related to matrix model by means of the Virasoro constraint and
string equation \cite{witten,Douglas,asv2}. Two sub-herarchies (BKP and CKP) of the KP hierarchy also  possess
the additional Virasoro symmetry \cite{DKJM-KPBKP,DJKM,kt1,heLMP} with consideration of
the reductions on the Lax operator.
The 2-dimensional Toda Lattice (2dTL) hierarchy introduced by Ueno
and Takasaki in \cite{takasaki1}  is
natural to have  the similar additional symmetry
because of the similarity between the KP hierarchy and this
hierarchy \cite{asv2}. For the dispersionless Toda
hiearchy \cite{Takasakicmp,Takasaki},  additional symmetries can be used to give
string equations and to solve Riemann-Hilbert problems.

Infinite dimensional Lie algebras of Block type,
as generalizations of the well-known Virasoro algebra,  have
been studied intensively in literature (see, e.g., \cite{Block,DZ,Su,Xu1}).  In \cite{ourBlock}, we provide  a Block type algebraic structure  of the bigraded Toda hierarchy (BTH) \cite{ourJMP,solutionBTH}.
Later on, this Block type Lie algebra is found again in dispersionless bigraded Toda
 hierarchy \cite{dispBTH}.
We would like to mention that this Block type Lie algebra is in fact also a special case of
Hamiltonian Lie algebras.
As one of four families of the well-known infinite dimensional Lie algebras of Cartan type (see, e.g., \cite{Kac,X}),
Hamiltonian Lie algebras, which
 usually possess additional structures of
associative algebras such that they form Poisson algebras,
 appear naturally in Hamiltonian mechanics, and are also central in the study of quantum groups.

For KP type integrable system, as  a reduction of two-component BKP hierarchy \cite{DJKM,takasakidBKP,LWZ,shiota,WX,TUdBKP}, the  Drinfeld-Sokolov hierarchy of D type \cite{LWZ,WX,DS,bkpds} has similar reduced double  dressing structures as BTH system and also has a Block (or Hamiltonian) type Lie symmetric structure \cite{blockBKPDS}.
 A nature question is whether we can find Block (or Hamiltonian) type Lie symmetry in these two kinds of dispersionless KP type integrable system (the dispersionless two-component BKP hierarchy and dispersionless D type Drinfeld-Sokolov hierarchy).
In this paper, the dispersionless D type Drinfeld-Sokolov hierarchy is proved to be a good model to derive
 Block  (or Hamiltonian) type infinite dimensional Lie symmetry.

This paper is arranged as follows. As a reduction of the dispersionless two-component BKP hierarchy,
the dispersionless D type Drinfeld-Sokolov hierarchy will be introduced in
Section 2. The Block  (or Hamiltonian) symmetries of the dispersionless D type Drinfeld-Sokolov hierarchy will be derived in Section 3.

\section{Dispersionless D type Drinfeld--Sokolov hierarchy}
\subsection{Dispersionless  two-component BKP hierarchy}

Underlying topological Landau-Ginzburg models of D-type, the dispersionless  two-component BKP hierarchy was proposed in \cite{takasakidBKP,TUdBKP}
\begin{align}\label{PPht}
& \frac{\pd L}{\pd t_k}=\{(L^{ k})_+, L\}, \quad \frac{\pd L}{\pd \hat t_k}=\{-(\hat{L}^{k})_-, L\},\ \ k\in\Zop,
\end{align}
\begin{align}\label{PPht}
& \frac{\pd \hat{L}}{\pd t_k}=\{(L^{ k})_+, \hat{L}\}, \quad \frac{\pd \hat{L}}{\pd \hat t_k}=\{-(\hat{L}^{k})_-, \hat{L}\},\ \ k\in\Zop,
\end{align}
where
\begin{align}&
 L=p+\sum_{i=1}^{\infty}\bar u_ip^{1-2i},\ \ \ \hat{L}=\sum_{i=1}^{\infty}\hat u_ip^{2i-1},
\end{align}
and the Poisson bracket $\{ \ ,\ \} $ is defined by
\begin{align}\{f, g\}:=ad\  f(g)=\frac{\partial f}{\partial p}\frac{\partial g}{\partial x}-\frac{\partial f}{\partial x}\frac{\partial g}{\partial p}.
\end{align}
It is easily to find the following antisymmetric condition
\begin{align}
 L(-p)=-L(p),\ \ \ \hat{L}(-p)=-\hat{L}(p).
\end{align}

\subsection{Dispersionless D type Drinfeld--Sokolov hierarchy}
Assume a new Lax function $\cL$ which has the following relation with two Lax functions of the dispersionless two-component BKP hierarchy introduced in last subsection
\begin{equation}\label{constraint}\cL=L^{2n}=\hat{L}^{-2}.\end{equation}
Then the Lax functions of dispersionless two-component BKP hierarchy will be reduced to the following Lax function of dispersionless D type Drinfeld--Sokolov hierarchy \cite{LWZ}
\begin{equation}\label{mL}
   \cL=p^{2n}+\sum_{i=1}^{n}u_i
p^{2i-2} +\rho^2p^{-2}.
\end{equation}

One can easily find the Lax function $\cL$ of  dispersionless D type Drinfeld--Sokolov hierarchy will satisfy the following symmetric condition
\begin{equation}
   \cL(-p)= \cL(p).
\end{equation}

The reduction eq.~\eqref{constraint} inspires us to define two fractional operators as
\begin{equation}
\cL^{\frac1{2n}}= p+\sum_{i\ge1}\bar u_i  p^{1-2i}, \quad \cL^{\frac12}=
\hat{u}_{-1}p^{-1}+\sum_{i\ge1}\hat{u}_i p^{2i-1},
\end{equation}
which satisfy
\begin{equation} \label{PPh}
\cL^{\frac1{2n}}(-p)=-\cL^{\frac1{2n}}(p),\quad \cL^{\frac12}(-p)=-\cL^{\frac12}(p).
\end{equation}

 The   dispersionless D type Drinfeld--Sokolov hierarchy being considered in this paper is defined by the following
Lax equations:
\begin{align}\label{PPht}
& \frac{\pd \cL}{\pd t_k}=\{(\cL^{\frac k{2n}})_+, \cL\}, \quad \frac{\pd \cL}{\pd \hat t_k}=\{-(\cL^{\frac k2})_-, \cL\},\ \ k\in\Zop.
\end{align}
One can find the terms with $p^0$ in $\cL^{\frac k{2n}}$ and $\cL^{\frac k{2}}$ disappear.

This Lax function $\cL$ of dispersionless D type Drinfeld--Sokolov hierarchy has the following dressing structure
  \begin{align}\label{dressing}\cL=e^{ad\varphi}(p^{2n} )= e^{ad \hat \varphi}(p^{-2}).\end{align}
The two dressing functions have the following form \begin{align}
 \varphi=\sum_{i=1}^{\infty}w_ip^{1-2i},\ \ \ \hat \varphi=\sum_{i=1}^{\infty}\hat w_ip^{2i-1}. \label{dressQ}
\end{align}

By tedious but standard computation, dispersionless Sato equations of dispersionless D type Drinfeld--Sokolov hierarchy can be derived in the following proposition.

\begin{prp}$\cL$ is the Lax function of the
dispersionless D type Drinfeld--Sokolov hierarchy if and only if  there exists two Laurent series $\varphi$
$\hat \varphi$ ({\it dressing fucntion}) which satisfy the equations
\begin{align}\label{sato dis}
    \nabla_{t_k,\varphi} \varphi = -(\cL^{\frac k{2n}})_-,\quad
     \nabla_{t_{k},\hat\varphi} \hat\varphi =  (\cL^{\frac{k}{2n}})_{\geq1},\ \ k\in\Zop,
\end{align}
\begin{align}
     \nabla_{\hat t_{k},\varphi} \varphi = -(\cL^{\frac{k}{2}})_-,\quad
     \nabla_{\hat t_{k},\hat\varphi} \hat\varphi = (\cL^{\frac{k}{2}})_{\geq1},\ \ k\in\Zop,
\end{align}
where $\nabla_{t,X} Y =\sum_{m=0}^{\infty}\frac{(ad X)^m}{(m+1)!} \partial_t Y.$

\end{prp}

In the next section, we shall show that this dispersionless D type Drinfeld-Sokolov hierarchies have a nice Block (or Hamiltonian) symmetry as its appearance  in BTH
\cite{ourBlock}.

\section{Block (or Hamiltonian) symmetries of dispersionless D type Drinfeld-Sokolov hierarchies }
\subsection{Block (or Hamiltonian) symmetries}
In this section, we will construct the flows of additional
symmetry which form the well-known Block (or Hamiltonian) type infinite dimensional Lie algebra.
Firstly we introduce dispersionless Orlov-Schulman operators as follows,
\begin{equation*}
\cM=e^{ad \varphi}(\Gm), \quad
\hat{\cM}=e^{ad \hat\varphi}(\hat\Gm),
\end{equation*}
where
\[
\Gm=\frac{xp^{1-2n}}{2n}+\sum_{k\in\Zop}\frac{k}{2n} t_k  p^{k-2n}, \quad \hat{\Gm}=-\frac x{2}p^{3}-\frac1{2}\sum_{k\in\Zop}
k\hat{t}_k p^{2-k}.
\]
We can rewrite $\cM,\hat{\cM}$ in terms of $\cL$ as
\begin{align}
\label{MintermL}
\cM=\frac{x\cL^{\frac{1}{2n}-1}}{2n}+\sum_{k\in\Zop}v_k \cL^{-\frac{k}{2n}-1}+\sum_{k\in\Zop}\frac{k}{2n} t_k  \cL^{\frac{k}{2n}-1},
\end{align}\begin{align}
\label{hMintermL}
 \hat{\cM}=-\frac x{2}\cL^{-\frac{3}{2}}+\sum_{k\in\Zop}\bar v_k \cL^{-\frac{k}{2}-1}-\frac1{2}\sum_{k\in\Zop}
k\hat{t}_k \cL^{\frac{k}{2}-1}.
\end{align}
It is easy to see the following lemma holds.
\begin{lem}\label{thm-Mw}
The operators $\cM$ and $\hat{\cM}$ satisfy
\begin{equation}\label{}
\{\cL, \cM\}=1, \quad \{\cL,\hat{\cM}\}=1;
\end{equation}
and
\begin{equation}\label{Mt}
\frac{\pd \bar\cM}{\pd t_k}=\{(\cL^{\frac k{2n}})_+,\bar\cM\},\quad
 \frac{\pd \bar\cM}{\pd \hat{t}_k}=\{-(\cL^{\frac k{2}})_-, \bar\cM\},
\end{equation}
where
 $\bar{\cM}=\cM$ or $\hat{\cM}, k\in\Zop$.
\end{lem}

\begin{prf}
The following calculation will lead to one part
of the first equation of eq.~\eqref{Mt}
\begin{align*}
&\partial_{ t_k}\cM=
\partial_{ t_k}e^{ad \varphi}(\Gamma)\\
&\phantom{\partial_{ t_k}\cM}
=e^{ad \varphi}\partial_{
t_k}(\Gamma)+\{\nabla_{t_k, \varphi}
\varphi,\cM \}\\
&\phantom{\partial_{ t_k}\cM}
= e^{ad \varphi}(\frac k{2n}
p^{k-2n})+\{-(\cL^{\frac k{2n}})_{-},\cM \}\\
&\phantom{\partial_{ t_k}\cM}
=\frac k{2n}\cL^{\frac k{2n}-1}+\{-(\cL^{\frac k{2n}})_{-},\cM \}\\
&\phantom{\partial_{ t_k}\cM}
=\{\cL^{\frac k{2n}},\cM\}+\{-(\cL^{\frac k{2n}})_{-},\cM \}\\
&\phantom{\partial_{ t_k}\cM}
= \{(\cL^{\frac k{2n}})_{+},\cM\}.
\end{align*}
The other parts can be proved in similar ways.
\end{prf}

We can formulate the following 2-form
\begin{align}\label{omega}
\omega&=d p\wedge dx+\sum_{k\in\Zop}d
(\cL^{\frac k{2n}})_+\wedge dt_{k}-\sum_{k\in\Zop}d
(\cL^{\frac k{2}})_-\wedge d\hat t_{k}
\end{align}
which satisfies
\begin{equation}
d\omega=0,\ \
\omega\wedge\omega=0.
\end{equation}
Further the following proposition can be derived.
\begin{prp}
 The dispersionless D type Drinfeld--Sokolov hierarchy is equivalent to the following exterior differential
equations.
\begin{equation}\label{wedge}
d\cL\wedge d\cM= d\cL\wedge d\hat \cM=\omega.
\end{equation}
\end{prp}
\begin{prf}
The proof is standard and one can check the similar proofs in\linebreak \cite{Takasaki,blockBKPDS}.
\end{prf}

One can  prove the following antisymmetric property of the dispersionless Orlov-Schulman operators $\cM$ and $\hat \cM$,
\begin{align}
\cM(-p)=-\cM(p),\ \ \ \hat \cM(-p)=-\hat \cM(p).
\end{align}

For dispersionless D-type Drinfeld-Sokolov hierarchy, we define
\begin{align}
\B_{m,l}=(\cM-\hat \cM)^{m}\cL^l,\ \  m\in \Zop,
\end{align}
and it is easy to check that
\begin{align}
\B_{m,l}(-p)=-\B_{m,l}(p),\ \  m\in \Zop.
\end{align}
That means it is reasonable to define additional flows of the dispersionless D type Drinfeld--Sokolov hierarchy as
\begin{align}\label{blockflow}
&\frac{\pd \cL}{\pd c_{m,l}}=\{-({\B}_{m,l})_-, \cL\},\ \ m \in \Zop, l\in \Z_+.
\end{align}
The action of above additional flows on $\varphi,\hat\varphi$ should be as the following
\begin{align}\label{addi-wave}
    \nabla_{c_{m,l},\varphi} \varphi& = -({\B}_{m,l})_-,\quad
     \nabla_{c_{m,l},\hat\varphi} \hat\varphi =  ({\B}_{m,l})_+.
\end{align}

Further we can get the following identities hold
\begin{equation}\label{blockflowM}
\frac{\pd \cM}{\pd c_{m,l}}\!=\!\{-(\B_{m,l})_-,\cM\}, \
\frac{\pd\hat{\cM}}{\pd c_{m,l}}\!=\!\{(\B_{m,l})_+,\hat{\cM}\},\   m\in \Zop, l\in \Z_+.\!\!\!\!\!\!\!\!
\end{equation}
The commutativity between different flows is a crucial property of integrable system. We shall show that  additional flows defined by  eq.~\eqref{blockflow} are commutative with flows of the dispersionless D type Drinfeld--Sokolov hierarchy (see the following proposition).
\begin{prp}\label{symmetry}
The additional flows in eq.~\eqref{blockflow} can commute with the original flow of  dispersionless Drinfeld--Sokolov hierarchy of type $D$, namely,
\begin{equation*}
\left[\frac{\pd}{\pd c_{m,l}}, \frac{\pd}{\pd t_k}\right]=0, \quad
\left[\frac{\pd}{\pd c_{m,l}}, \frac{\pd}{\pd\hat{t}_k}\right]=0, \qquad m,k\in\Zop, l
\in\Z_+
\end{equation*}
which hold in the sense of acting on  $\cL.$

\end{prp}
\begin{prf} According to the definition, direct calculations can lead to
\begin{eqnarray*}
[\partial_{c_{m,l}},\partial_{t_k}]\cL&=&\partial_{c_{m,l}}
(\partial_{t_k}\cL)-
\partial_{t_k} (\partial_{c_{m,l}}\cL),\\
&=& -\partial_{c_{m,l}}\{(\cL^{\frac{k}{2n}})_{-},\cL\}+
\partial_{t_k} \{((\cM-\hat \cM)^m\cL^l)_{-},\cL \}\\
&=& \{-(\partial_{c_{m,l}}\cL^{\frac{k}{2n}} )_{-},\cL\}-
\{(\cL^{\frac{k}{2n}})_{-},(\partial_{c_{m,l}}\cL)\}\\&&+
\{[\partial_{t_k} ((\cM-\hat \cM)^m\cL^l)]_{-},\cL\} +
\{((\cM-\hat \cM)^m\cL^l)_{-},(\partial_{t_k}\cL)\}.
\end{eqnarray*}
Using eq.~\eqref{PPht} and eq.~\eqref{Mt}, it
equals
\begin{eqnarray*}
[\partial_{c_{m,l}},\partial_{t_k}]\cL
&\!\!\!=\!\!\!&\{\{((\cM-\hat \cM)^m\cL^l)_{-}, \cL^{\frac{k}{2n}}\}_{-},\cL\}\!+\!
\{(\cL^{\frac{k}{2n}})_{-},\{((\cM-\hat \cM)^m\cL^l)_{-},\cL\}\}\\
&\!\!\!\!\!\!\!&\!+\{\{(\cL^{\frac{k}{2n}})_{+},(\cM\!-\!\hat \cM)^m\cL^l\}_{-},\cL\}\!-\!\{((\cM\!-\!\hat \cM)^m\cL^l)_{-},\{(\cL^{\frac{k}{2n}})_{-},\cL\}\}\\
&\!\!\!=\!\!\!&\{\{((\cM-\hat \cM)^m\cL^l)_{-}, \cL^{\frac{k}{2n}}\}_{-},\cL\}- \{\{(\cM-\hat \cM)^m\cL^l,
(\cL^{\frac{k}{2n}})_{+}\}_{-},\cL\}\\&\!\!\!\!\!\!&\!+
\{\{(\cL^{\frac{k}{2n}})_{-},((\cM-\hat \cM)^m\cL^l)_{-}\},\cL\}\\
&\!\!=\!&0.
\end{eqnarray*}
The other cases of this proposition can be proved in similar ways.
\end{prf}

Proposition \ref{symmetry} indicates that eq.~\eqref{blockflow} is a symmetry of dispersionless D type Drinfeld-Sokolov hierarchy. Thus the flows in eq.~\eqref{blockflow} are called additional symmetry flows of the dispersionless D type
  Drinfeld-Sokolov hierarchy.

In order to have a better understanding of the properties
of the additional symmetry flows, it is necessary to determine their algebraic structure.
Using same techniques as in \cite{dispBTH}, the following theorem can be derived.
\begin{thm}\label{thm-MLs}
The flows in eq.~\eqref{blockflow} about  additional symmetries of  dispersionless D type Drinfeld-Sokolov hierarchy compose the following Block $($or Hamiltonian$)$ type Lie algebra
\begin{eqnarray}\label{blockrelation}
[\partial_{c_{m,l}},\partial_{c_{s,k}}]=(km-s l)\partial_{c_{m+s-1,k+l-1}},\ \  m,s \in \Zop,\,k,l\in \Z_+,
\end{eqnarray}
which holds in the sense of acting on  $\cL.$
\end{thm}
\begin{prf}
 By using eq.~\eqref{blockflow} and eq.~\eqref{blockflowM}, we get
\begin{eqnarray*}
[\partial_{c_{m,l}},\partial_{c_{s,k}}]\cL&=&
\partial_{c_{m,l}}(\partial_{c_{s,k}}\cL)-
\partial_{c_{s,k}}(\partial_{c_{m,l}}\cL)\\
&=&-\partial_{c_{m,l}}\{((\cM-\hat \cM)^s\cL^k)_{-},\cL\}
+\partial_{c_{s,k}}\{((\cM-\hat \cM)^m\cL^l)_{-},\cL\}\\
&=&-\{(\partial_{c_{m,l}}
(\cM-\hat \cM)^s\cL^k)_{-},\cL\}-\{((\cM-\hat \cM)^s\cL^k)_{-},(\partial_{c_{m,l}} \cL)\}\\
&&+ \{(\partial_{c_{s,k}} (\cM-\hat \cM)^m\cL^l)_{-},\cL\}+
\{((\cM-\hat \cM)^m\cL^l)_{-},(\partial_{c_{s,k}} \cL)\},
\end{eqnarray*}
which further leads to
 \begin{eqnarray*}&&
[\partial_{c_{m,l}},\partial_{c_{s,k}}]\cL\\
&=&-\{\Big[(\partial_{c_{m,l}}(\cM-\hat \cM))(\cM-\hat \cM)^{s-1}\cL^k
+(\cM-\hat \cM)^s(\partial_{c_{m,l}}\cL^k)\Big]_{-},\cL\}\\&&-\{((\cM-\hat \cM)^s\cL^k)_{-},(\partial_{c_{m,l}} \cL)\}\\
&&+\{\Big[(\partial_{c_{s,k}}(\cM-\hat \cM))(\cM-\hat \cM)^{m-1}\cL^l
+(\cM-\hat \cM)^m(\partial_{c_{s,k}}\cL^l)\Big]_{-},\cL\}\\&&+
\{((\cM-\hat \cM)^m\cL^l)_{-},(\partial_{c_{s,k}} \cL)\}\\
&=&\{[(s l-km)(\cM-\hat \cM)^{m+s-1}\cL^{k+l-1}]_-,\cL\}\\
&=&(km-s l)\partial_{c_{m+s-1,k+l-1}}\cL.
\end{eqnarray*}
\end{prf}

As indices $ m,s$ in eq.~\eqref{blockrelation} can only take values in odd numbers, the algebra is  a subalgebra of the  Lie algebra considered in \cite{ourBlock}, which is a Block type  Lie algebra as well as a special case of Hamiltonian type Lie algebras (see, e.g., \cite{Kac,X}).
The above results together with that on  \cite{dispBTH} shows that the Block (or Hamiltonian) type Lie algebras
appear not only in  dispersionless Toda type  systems (dispersionless bigraded Toda hierarchy)
  but also in  dispersionless  KP type integrable systems (dispersionless D type
  Drinfeld-Sokolov hierarchy). These results also provide a strong support of the universality
of Block (or Hamiltonian) type infinite dimensional Lie algebra  in integrable hierarchies.

To see these Block (or Hamiltonian) symmetry flows more clearly, we give one example about a specific Block (or Hamiltonian) symmetry flow equation in the next subsection.

\subsection{Example of Block  (or Hamiltonian) symmetry flow}
Here we consider the dispersionless D type
  Drinfeld-Sokolov hierarchy when $n=1$, i.e. the hierarchy with Lax function as   \begin{equation}\label{mLn=1}
   \cL=p^{2}+u+\rho^2p^{-2}.
\end{equation}
As the first simplest nontrivial example, we shall discuss Block (or Hamiltonian) symmetry flow when $m=l=1$ in eq.~\eqref{blockflow}, i.e.
\begin{align}\label{backlund}
&\frac{\pd \cL}{\pd c_{1,1}}=\{-({\B}_{1,1})_-, \cL\}=\{-((\cM-\hat \cM)\cL)_-, \cL\}.
\end{align}
To avoid confusion, we denote different fractional functions $(\cL^{ \frac{ k}{2n}},\cL^{ \frac{ k}{2}})$ when $n=1$ as $(\cL^{ \frac{ k}{2}},\cL^{ \frac{\hat k}{2}})$ respectively.
From eq.~\eqref{MintermL} and eq.~\eqref{hMintermL}, one can easily get
\begin{eqnarray*}\label{MintermL2}
((\cM-\hat \cM)\cL)_-&=&\frac{x\cL^{\frac{1}{2}}}{2}+\sum_{k\in\Zop}v_k \cL^{-\frac{k}{2}}+\sum_{k\in\Zop}\frac{k}{2} t_k  \cL^{\frac{k}{2}}+\frac1{2}\sum_{k\in\Zop}
k\hat{t}_k \cL^{ \frac{\hat k}{2}}.
\end{eqnarray*}

From dressing structure eq.~\eqref{dressing} and eq.~\eqref{MintermL}, the following relations between $u,\rho$ and $v_1,v_3$ can be derived
 \begin{equation}\label{mL2}
  u=-2\omega_{1x}=4v_{1x},\ \ v_3=-\frac{\omega_{1}\omega_{1x}}{4}-\frac32\omega_3,\ \ \rho^2=\omega_{1}\omega_{1x}-2\omega_3.
\end{equation}
Suppose $\cL$ only depends on $x,t_1,\hat t_1,c_{1,1}$, then we can get
\begin{align}\notag
\frac{\pd \cL}{\pd c_{1,1}}&=\{-\frac{(x+ t_1)\cL_-^{\frac{1}{2}}}{2}-v_1 \cL^{-\frac{1}{2}}-v_3 \cL^{-\frac{3}{2}}-\frac1{2}\hat{t}_1 \cL_-^{ \frac{\hat 1}{2}}, \cL\}\\ \notag
&=\{-\frac{(x+ t_1)\cL_-^{\frac{1}{2}}}{2}-v_1 \cL^{-\frac{1}{2}}-v_3 \cL^{-\frac{3}{2}}-\frac1{2}\hat{t}_1 \rho p^{-1}, \cL\}\\
&=\frac{u}{2}+\frac{(x+ t_1)}{2}\partial_{t_1}u+ 2v_{1x}+\hat{t}_1 \rho_x\\ \notag
&\ \ +(\frac{(x+ t_1)}{2}\partial_{t_1}(\rho^2)+ 2v_{3x}+\frac1{2}\hat{t}_1 \rho u_{x})p^{-2}.
\end{align}
which further leads to the following additional coupled flow equations over dynamic variables $u,\rho$
\begin{align}\label{backlund}
\frac{\pd u}{\pd c_{1,1}}&=u+\frac{(x+ t_1)}{2}\partial_{t_1}u+\hat{t}_1\rho_x,\\
\frac{\pd \rho}{\pd c_{1,1}}&=\frac{3\rho_x}{2}+\frac{(x+ t_1)}{2}\partial_{t_1}\rho+\frac14(u^2+u_{x}\int u dx)\rho^{-1}+\frac1{4}\hat{t}_1u_{x}.
\end{align}
\vskip 0.5truecm {\small \noindent{\bf Acknowledgments.}
Chuanzhong Li is supported by the National Natural Science Foundation of China under Grant No. 11201251,
the Natural Science Foundation of Zhejiang Province under Grant No. LY12A01007 and the Natural Science Foundation of Ningbo under Grant No. 2013A610105.
Jingsong He is supported by the National Natural Science Foundation of China under Grant No. 11271210 and K.C.Wong Magna Fund in
Ningbo University. Yucai Su is supported by the National Science Foundation of China under Grant No. 11371278, the Shanghai Municipal Science and Technology Commission under Grant No. 12XD1405000 and the Fundamental Research Funds for the Central Universities of China.
}

\end{document}